\documentclass[onecolumn,showpacs,amsmath,amssymb,10pt,aps,nofootinbib]{revtex4}
\usepackage{amsmath}
\usepackage{amsfonts}
\usepackage{amssymb}
\newcommand{\p}{\partial}
\newcommand{\reff}[1]{(\ref{#1})}

\begin{document}

\title{On the Gravitational Collapse of a Gas Cloud in Presence of Bulk Viscosity}

\author{Nakia Carlevaro and Giovanni Montani}
    \affiliation{ICRA--International Center for Relativistic Astrophysics c/o Dipartimento di Fisica (G9), Universit\`a       di Roma ``La Sapienza'', Piazza A.Moro 5, 00185 Rome, Italy \\and \\Dipartimento di Fisica Universit\`a di Roma          ``La Sapienza'' \\nakia.carlevaro@icra.it\\montani@icra.it}

\date{\today}

\begin{abstract}
We analyze the effects induced by the \emph{bulk} (or second) \emph{viscosity} on the dynamics associated to the extreme gravitational collapse. Aim of the work is to investigate whether the presence of viscous corrections to the evolution of a collapsing gas cloud influence the \emph{top-down} fragmentation process. To this end, we generalize the approach presented in the Hunter work \cite{h62} to include in the dynamics of the (uniform and spherically symmetric) cloud the negative pressure contribution associated to the bulk viscosity phenomenology. Within the framework of a Newtonian approach (whose range of validity is outlined), we extend to the viscous case either the Lagrangian, either the Eulerian motion of the system addressed in \cite{h62} and we treat the asymptotic evolution in correspondence to a viscosity coefficient of the form $\zeta=\zeta_0\,\rho^{5/6}$ ($\rho$ being the cloud density and $\zeta_0=const.$).

We show how the adiabatic-like behavior of the gas is deeply influenced by viscous correction when its collapse reaches the extreme regime toward the singularity. In fact the density contrast associated to a given scale of the fragmentation process acquires, asymptotically, a vanishing behavior which prevents the formation of sub-structures. Since in the non-viscous case the density contrasts remain constant, we can conclude that in the adiabatic-like collapse the \emph{top down} mechanism of structures formation is suppressed as soon as viscous effects are taken into account. Such a feature is not present in the isothermal-like collapse because the sub-structures formation is yet present and outlines the same behavior as in the non-viscous case. We emphasize that in the adiabatic-like collapse the \emph{bulk viscosity} is also responsible for the appearance of a threshold scale beyond which perturbations begin to increase; this issue, absent in the non-viscous case, is equivalent to deal with a Jeans length. A discussion of the physical character that the choice $\nu=5/6$ takes place in the present case is provided. 
\end{abstract}

\pacs{95.30.Wi, 51.20.+d}
\maketitle
\section{Introduction}
One of the most attractive challenge of Relativistic Cosmology is to provide a self-consistent picture for processes of gravitational instability which connect the high isotropy of the cosmic microwaves background radiation \cite{h03}, \cite{p03} with the striking clumpiness of the actual Universe \cite{t04}. An interesting framework is provided by the \emph{top-down} scheme of structure fragmentation, mainly associated with the hot dark matter phenomenology \cite{be80}, \cite{bs83}, \cite{max03}. Such a scheme is based on the idea that perturbation scales, contained within a collapsing gas cloud, start to collapse (forming sub-structures) because their mass overcomes the decreasing Jeans value \cite{j902}, \cite{j28}, \cite{b56}, \cite{b57} of the background system. The resulting effect of such a gravitational instability consists of a progressive enhancement of the density contrasts associated to the perturbations sub-scales.

In a work by C. Hunter \cite{h62}, a specific model for a gas cloud fragmentation was addressed and the behavior of sub-scales density perturbations, outcoming in the extreme collapse, was analytically described. The hypotheses on which this model is based are the homogeneity and the spherical symmetry, respectively, of the collapsing cloud that starts at rest its fall. Furthermore, it is assumed that pressure forces are negligible in the collapse dynamics and therefore a real notion of Jeans mass is not required in this approach. According to this scheme, in \cite{h62} is developed the Lagrangian and Eulerian formulation of the zero and first-order dynamics respectively. The results of this analysis show that density contrasts grow approaching the singularity inducing a fragmentation process of the basic flow. It is outlined that pressure effects do not influence the perturbations behavior if we consider an isothermal-like politropic index $\gamma$ (i.e. for $1\leqslant\gamma<4/3$). On the other hand, such effects increase as $\gamma$ runs from $4/3$ to the adiabatic value $5/3$. Just this limit case, for $\gamma=5/3$, represents an exception being characterized by density contrasts which remain constant asymptotically to the singularity. 

In this work we investigate how the above picture is modified by including, in the gas cloud dynamics, the presence of \emph{bulk viscosity} effects. In this respect, we generalize the Lagrangian evolution by taking into account the force acting on the collapsing shell as a result of the negative pressure connected to the presence of \emph{bulk viscosity}. We construct such an extension requiring that the asymptotic dynamics of the collapsing cloud is not qualitatively affected by the presence of viscosity. In particular, we analytically integrate the dynamics in correspondence to the constitutive equation \cite{bk76}, \cite{bk77}, \cite{bnk79}, \cite{m95} for the viscosity coefficient $\zeta=\zeta_0\,\rho^{\nu}$ where the exponent is assumed to be $\nu=5/6$ ($\rho$ being the cloud unperturbed density and $\zeta_0$ a constant parameter). Then we face the Eulerian motion of the inhomogeneous perturbations living within the cloud. The resulting viscous dynamics is treated in the asymptotic limit to the singularity. We show that density contrasts behave, in the isothermal-like collapse, as in the non-viscous case. On the other hand, the perturbations damping increases monotonically as $\gamma$ runs from $4/3$ to $5/3$; in fact, for such adiabatic-like case, we see that density contrasts asymptotically vanish and no fragmentation processes take place in the cloud when the viscous corrections are sufficiently large. In particular, we observe the appearance of a threshold value for the scale of the collapsing perturbations depending on the values taken by the parameters $\zeta_0$ and $\gamma\in(4/3,5/3]$; such a viscous effect corresponds to deal with an analogous of the Jeans length, above which perturbations are able to collapse. However such a threshold value does not ensure the diverging behavior of density contrasts which takes place, in turn, only when a second (greater) critical length is overcome.

Since in the extreme collapse it is expected that viscous processes are relevant, our analysis suggests that the \emph{top-down} scheme of structures formation can be deeply influenced when non-equilibrium features of the dynamics arise. Indeed \emph{bulk viscosity} outcomes as a phenomenological issue of a thermodynamical non-equilibrium collapse and its presence has to be consider important in the non-linear regime of structures formation. According to our analysis, if such viscous effects are sufficiently intense, the final system configuration is not a fragmented cloud as a cluster of sub-structures but simply a single object (a black hole, in the present case, because pressure forces are taken negligible). Furthermore, we discuss why the choice $\nu=5/6$ has a physical meaning in the viscous dynamics: we show that for $\nu>5/6$ the background evolution would be asymptotically affected by viscosity which would acquire a non-perturbative character; on the other hand, for $\nu<5/6$ no modifications occur with respect to the dynamics of the non-viscous density contrasts. 

The scheme of the paper is as follow. In Sec. \ref{Sec.2} a review of the Hunter work \cite{h62} is presented; after having integrated the basic flow motion equation we analyze the asymptotic (toward the singularity) evolution of the density perturbations in the cases of an isothermal and adiabatic behavior of the gas respectively. In Sec. \ref{Sec.3} we discuss how the presence of \emph{bulk viscosity} influences the zero-order gas cloud dynamics which is supposed to be qualitatively the same as in the non-dissipative case. The analysis of the Lagrangian motion is developed for a particular value of the \emph{bulk viscosity} power-law exponent $\nu=5/6$. In Sec. \ref{Sec.4} we start from the unperturbed solutions of the background field to develop a perturbation theory in a Newtonian approach. The Eulerian motion equations are written for first-order perturbed quantities and then combined to build an unique second-order differential equation which governs the density perturbations evolution. In Sec. \ref{Sec.5} we integrate this fundamental equation to obtain the asymptotic behavior of the fluctuations for the final part of the collapse. Two different cases are analyzed: the isothermal-like one which stands for $1\leqslant\gamma<4/3$ and the adiabatic-like case for $4/3<\gamma\leqslant5/3$. At the end of this Section are pointed out the physical reasons to have a viscous fluid with $\nu=5/6$ and the conditions for the validity of the Newtonian approximation. In Sec. \ref{Sec.6} brief concluding remarks follow and possible upgradings for the description of the \emph{bulk viscosity} effects on the gravitational collapse are discussed.

\section {Description of a non-viscous cloud fragmentation}\label{Sec.2}
In this Section we present an hydrodynamical analysis of a spherically symmetric gas cloud collapse. This model was firstly proposed by Christopher Hunter in 1962 \cite{h62}, in his work he supposed that the gas cloud becomes unstable with respect to its own gravitation and begin to condense. The collapsing cloud is assumed to be the dynamical background on which studying, in a Newtonian regime, the evolution of density perturbations generated on this basic flow. Such an analysis is suitable for the investigation of the cosmological structures formation in the \emph{top-down} scheme \cite{be80}, \cite{bs83} since it deals with the sub-structures temporal evolution compared with the basic flow of the gravitational collapse.

In the Hunter model, the unperturbed flow was supposed to be homogeneous, spherically symmetric and initially at rest. Furthermore the gravitational forces are assumed to be very much greater than the pressure ones, which are therefore neglected in the zero-order analysis. In such an approach the gas results to be unstable, since there are no forces which can contrast the collapse, and the condensation starts immediately.

The basic flow is governed by the Lagrangian motion equation of a spherically symmetric gas distribution which collapses under the only gravitational action. Assuming that the initial density of the cloud is constant in space, the dynamics reads
\begin{equation}
\label{eq-lagrangian-0}
\frac{\p^2r}{\p t^2}=-\frac{GM}{r^2}\;,
\end{equation}
where the origin $O$ is taken at the center of the gas, $r$ is the radial distance, $G$ the gravitational constant and $M$ the mass of the gas inside a sphere of radius $r$. In what follows, we shall suppose that the gas was at a distance $a$ from $O$ in correspondence to the initial instant $t_0$; this distance $a$ identifies a fluid particle and will be used as a Lagrangian independent variable so $r=r(a,t)$. Provided that particles do not pass trough each other, the mass $M$ inside a sphere of radius $r$ is not time dependent and is a function of $a$ only; using the integral form
$M(a)=\int_0 ^r dr^{\prime}\, 4\pi\rho r^{\prime2} =\int_0 ^a da^{\prime}\, 4\pi\rho_0(a')a^{\prime2}$, where $\rho$ is the gas cloud density and $\rho_0=\rho(t_0)$ the initial one, we get the relation
\begin{equation}
\label{mass-conservation}
\rho r^2 \frac{\partial r}{\partial a}=\rho_0 a^2\; .
\end{equation}

A first integration of \reff{eq-lagrangian-0} yields the expression of the radial velocity $v=\p r/\p t$ which reads
\begin{equation}
\label{vel-v-0}
v=-[2GM\left(1/r-1/a\right)]^{\frac{1}{2}}\;,
\end{equation}
where we considered the negative solution since the radial unperturbed velocity must be negative in order to obtain a collapse. Let us now introduce the parametrization 
\begin{equation}
\label{r-a}
r=a\cos^2\beta\;,
\end{equation}
where $\beta=\beta(t)$ is a time dependent function such that $\beta(t_0)=0$ and $\beta(0)=\pi/2$ since we choose the origin of time to have $t=0$ when $r=0$ and $t_0$ takes negative values. Since we assumed $\rho_0$ to be uniform, we are now able to integrate equation \reff{vel-v-0} to get the following relation between $\beta$ and $t$ and the expression of the initial time $t_0$:
\begin{equation}
\label{beta-t-0}
\beta +\frac{1}{2}\sin 2\beta=\frac{\pi}{2}+t\sqrt{\frac{8}{3}\pi\rho_0G}\;,
\end{equation}
\begin{equation}
\label{t_0-0}
t_0=-\sqrt{3\pi\,/\,32\rho_0G}\;.	
\end{equation}
It is more convenient to use an Eulerian representation of the flow field. To this end, we now derivate the relation \reff{r-a} and integrate equation \reff{mass-conservation} to obtain respectively the unperturbed radial velocity $v$ and the basic flow density $\rho$. Furthermore, solving the well known Poisson equation for the gravitational potential $\phi$, we obtain the unperturbed solutions describing the background motion; all these quantities take the explicit forms
\begin{subequations}
\label{basic-flow}
\begin{align}
\label{V-basic-flow}
& \textbf{v}=[v,0,0]\;,\\
\label{v-basic-flow}
& v=-2r\dot{\beta}\tan\beta\;,\\
\label{rho-basic-flow}
& \rho=\rho_0\cos^{-6}\beta\;,\\
\label{phi-basic-flow}
& \phi=-2\pi\rho_0 G\left(a^2-r^2/3\right)\cos^{-6}\beta\;,
\end{align}
\end{subequations}
where ($\,\dot{}\,$) denotes the derivate with respect to time and in \reff{V-basic-flow} the non-radial components of velocity must vanish since we are considering a spherical symmetry.  

The first-order perturbations to the basic flow (higher orders analysis was made by Hunter in two later articles \cite{h64}, \cite{h77}) are investigated in the Newtonian limit starting from the continuity, Euler and Poisson equations \cite{ll-fluid}. In such a picture, we are now interested to study the effects of the thermostatic pressure on the perturbations evolution. We shall therefore consider terms due to the pressure $p$ in the motion equations, which read 
\begin{subequations}
\label{eq-unpert-0}
\begin{align}
\label{continuity-unpert-0}
&\dot{\rho}+\nabla\cdot(\rho\,\textbf{v})=0\;,\\
\label{eulero-unpert-0}
&\dot{\textbf{v}}+(\textbf{v}\cdot\nabla)\cdot\textbf{v}=-\nabla\phi-\frac{1}{\rho}\,\nabla p\;,\\
\label{poisson-unpert-0}
&\nabla^2\phi=4\pi G\rho\;. 
\end{align}
\end{subequations}
The gas is furthermore assumed to be barotropic, i.e. the pressure depends only by the background density $\rho$. In this model, zero-order solutions \reff{basic-flow} are already verified since the pressure gradient, in the homogeneity hypothesis, vanishes and the pressure affects only the perturbative dynamics.
  
Let us now investigate first-order fluctuations around the zero-order solutions, i.e. we take the perturbed quantities: $(\textbf{v}+\delta\textbf{v})$, $(\rho+\delta\rho)$, $(\phi+\delta\phi)$ and also $(p+\delta p)$ where $p=p\,(\rho)$. Substituting these solutions in the Euler equation \reff{eulero-unpert-0} and taking the $\textbf{rot}$ of the final expression, we get, linearizing in the perturbed quantities, an equation for the vorticity $\delta\textbf{w}=\nabla\times\delta\textbf{v}$ which stands
\begin{equation}
\label{eq-vorticity}
\dot{\delta\textbf{w}}=-\nabla\times\left[\delta\textbf{w}\times\textbf{v}\right]\;.
\end{equation}
Using spherical coordinates $[r(a,t), \theta, \varphi]$ we are able to build the solutions for the three components of the vorticity, getting
\begin{equation}
\label{dw}
\delta\textbf{w}=\left[\;l\;\cos^{-4}\beta +h\;,\quad m\;\cos^{-4}\beta\;,\quad n\;\cos^{-4}\beta\;\right]\;.		
\end{equation}
Here $l, m, n$ are arbitrary functions of the new variables $[a, \theta, \varphi]$ (the radial coordinate transforms like \reff{r-a}) which must satisfy the relation $\nabla \cdot \delta\textbf{w}=0$ \footnote{We remember that in any coordinates system the relation $div\,\textbf{rot}\equiv0$ stands.} and $h$ is physically irrelevant since it represents a static distribution of the $\delta\textbf{w}$ first component in the space.

We are now able to find a solution for the perturbed velocity using the vorticity expression; one can always consider a solution of the form 
\begin{equation}
\label{dv}
\delta\textbf{v}=\textbf{V}\cos^{-2}\beta+\nabla\psi\;,
\end{equation}
where $\psi$, $\textbf{V}$ are arbitrary functions of the coordinates and for $\textbf{V}$ we assume a restriction gives by the relation $\nabla\cdot\textbf{V}=0$.

Let us now write three equations for the perturbed quantities $\delta\textbf{v}$, $\delta\rho$ and $\delta\phi$. Substituting the last expression for the velocity fluctuation in equations \reff{eq-unpert-0} as perturbed to the first-order and eliminating the variable $r$ trough the relation \reff{r-a} we get the following equations
\begin{subequations}
\label{perturbed-eq-0}
\begin{align}
\label{pert-1-0}
&\dot{\delta\rho}-6\,\delta\rho\,\dot{\beta}\tan\beta+\rho_0\cos^{-10}\beta\;D^2\psi=0\;,\\
\label{pert-2-0}
&\dot{\psi}+\delta\phi+\frac{v_s^2}{\rho_0}\cos^6\beta\,\delta\rho=0\;,\\
\label{pert-3-0}
&D^2\delta\phi-4\pi G \cos^4\beta\,\delta\rho=0\;.
\end{align}
\end{subequations}
Here time differentiation is taken at some fixed comoving radial coordinate, $v_s$ is the sound speed given by $v_s= \sqrt{\p p/\p\rho}$ and $D^2$ is the Laplace operator as written in our comoving coordinates.

A single second-order differential equation for $\delta\rho$ can be obtained from the equations set \reff{perturbed-eq-0} by eliminating $\delta\phi$ between \reff{pert-2-0} and \reff{pert-3-0} and then eliminating $\psi$ between the resulting relation and equation \reff{pert-1-0}. This final equation is as follow
\begin{equation}
\label{eq-fundamental-0}
\tfrac{\p}{\p t}\left(\cos^{10}\beta\;\dot{\delta\rho}-6\;\sin\beta\,\cos^9\beta\,\dot{\beta}\,\delta\rho\right)-4\pi G\rho_{0}\,\cos^4\beta\,\delta\rho={v_s}^2\,\cos^6\beta\;D^2\delta\rho\;.
\end{equation}

In order to study the temporal evolution of density perturbations, we assume to expand $\delta\rho$ in plane waves of the form
\begin{equation}
\label{fourier-trasform}
\delta\rho(\textbf{r},t)=\delta\varrho\,(t)\;e^{-i\textbf{k}\cdot\textbf{r}}\;,
\end{equation}
where $1/k$ (with $k=|\textbf{k}|$) represents the initial length scale of the considered fluctuation. We shall now express the thermostatic pressure as a function of the basic flow density using the barotropical law
\begin{equation}
\label{barotropic}
p=\kappa\,\rho^\gamma\;,
\end{equation}
where $\kappa$, $\gamma$ are constants and $1\leqslant\gamma\leqslant5/3$. By this expression we are able to distinguish a set of different cases related to different values of the politropic index $\gamma$. The asymptotic value $\gamma=1$ represents an isothermal behavior of the gas cloud and corresponds to a constant sound speed $v_s$; the case $\gamma=5/3$ describes, instead, an adiabatic behavior and it will be valid when changes are taking place so fast that no heat is transferred between elements of the gas. We can suppose that intermediate values of $\gamma$ will describe intermediate types of behavior between the isothermal and adiabatic ones.   

The temporal evolution of density perturbations is governed by \reff{eq-fundamental-0}; this equation can not be solved in general but we can determine the asymptotic behavior of solutions for the final part of the collapse as $(-t)\rightarrow0$. In this limit, we are able to develop up to the first-order the equation \reff{beta-t-0} which, once integrated, gives the time dependence of the parameter $\beta$. For small $\beta$ we are able to approximate $\sin\beta\approx1$ in order to obtain the relation
\begin{equation}
\cos\beta^3=\sqrt{6 \pi G \rho_0}\;(-t)\;.	
\end{equation}
In this approach we determine the asymptotic temporal evolution of the basic flow unperturbed density \reff{rho-basic-flow} which now reads
\begin{equation}
\label{rho-t-basic-flow}
\rho\sim(-t)^{-2}\;.
\end{equation}
Substituting this expression in \reff{eq-fundamental-0}, together with \reff{fourier-trasform} and \reff{barotropic} we get the following asymptotic equation for the final part of the collapse
\begin{equation}
\label{eq-asym-0}
(-t)^2\,\ddot{\delta\varrho}-\frac{16}{3}(-t)\,\dot{\delta\varrho}+
\left[4+\frac{v_0^2\, k^2\,(-t)^{8/3-2\gamma}}{(6\pi G\rho_0)^{\gamma-1/3}}\right] \delta\varrho=0\;,	
\end{equation}
where $v_0^2=\kappa\gamma\rho^{\gamma-1}_{0}$. A complete solution of this equation involves Bessel functions and reads
\begin{equation}
\label{drho-solution-0}
\delta\varrho=(-t)^{-13/6}\;\Big[C_1\,J_n\big[q(-t)^{4/3-\gamma}\big]\,+C_2\,Y_n\big[q(-t)^{4/3-\gamma}\big]\Big]\;,
\end{equation}
where the parameters $n$ and $q$ are
\begin{subequations}
\label{bessel-parameter-0}
\begin{align}
\label{parameter-n-0}
&n=5\;/\,6(4/3-\gamma)\;,\\
\label{parameter-q-0}
&q=-v_0 k(6\pi \rho_0 G)^{1/6-\gamma/2}\;/\,(4/3-\gamma)\;.
\end{align}
\end{subequations}

In order to study the asymptotic evolution of this solution, we shall analyze the cases $1\leqslant\gamma<4/3$ and $4/3<\gamma\leqslant5/3$ separately, since Bessel functions have different limits connected to the magnitude of their argument \cite{jj}. In the asymptotic limit to the singularity, the isothermal-like case is characterized by a positive time exponent inside Bessel functions so $qt^{4/3-\gamma}\ll1$, on the other hand, in the adiabatic-like behavior we obtain $qt^{4/3-\gamma}\gg1$.
 
For $1\leqslant\gamma<4/3$ Bessel functions $J$ and $Y$ behave like a power-law of the form {\footnotesize{$J_n(x)\sim x^{+n}$}}, {\footnotesize{$Y_n(x)\sim x^{-n}$}}, for $x\ll1$. By this approximation, the solution \reff{drho-solution-0} assumes the the following asymptotic form
\begin{equation}
\label{rho-isoterm-0}
\delta\varrho^{\,ISO}\sim(-t)^{-3}\;,
\end{equation}
which holds for all the isothermal-like $\gamma$ values. This result implies that density perturbations grow to infinity as $(-t)\rightarrow0$; this behavior can be deducted by simply analyzing equation \reff{eq-fundamental-0}. Its right hand side contribution, due to pressure forces, is proportionally to $\cos^6\beta$ which decreases drastically in the last part of the collapse (i.e. $\beta\rightarrow\pi/2$), in this way pressure forces become negligible toward the condensation in comparison with respect to the gravitational ones and can not, in any cases, prevent the collapse of density fluctuations. Let us now study the asymptotic behavior of the density contrasts $\delta=\delta\varrho/\rho$. It is immediate to see that for all the values of $\gamma$ inside the interval $[1,4/3)$ density contrasts asymptotically diverge like
\begin{equation}
\label{delta-iso-0}
\delta^{\,ISO}\sim(-t)^{-1}\;,
\end{equation}
implying that perturbations grow more rapidly with respect to the back-ground density favoring the fragmentation of the basic structure independently on the value of the politropic index. 

In the adiabatic-like case, for $4/3<\gamma\leqslant5/3$, the argument of Bessel functions becomes much gather than unity and they assume an oscillating behavior like {\footnotesize{$J_n(x)\sim x^{-1/2}\cos(x)$}}, {\footnotesize{$Y_n(x)\sim x^{-1/2}\sin(x)$}}, for $x\gg1$. The solution \reff{drho-solution-0} asymptotically reads
\begin{equation}
\label{rho-adiabatic-0}
\delta\varrho^{\,ADB} \sim (-t)^{\gamma/2-17/6}\;\,^{\cos}_{\sin}\left[ \frac{v_0 k (-t)^{4/3-\gamma}}{(4/3-\gamma)(6\pi G\rho_0)^{\gamma/2-1/6}}\right]\;,
\end{equation} 
and therefore perturbations oscillate with ever increasing frequency and amplitude. In this case, density contrasts assume the form
\begin{equation}
\label{delta-adb-0}
\delta^{\,ADB}\sim(-t)^{\frac{\gamma}{2}-\frac{5}{6}}\;,
\end{equation}
and they outline that perturbations, for intermediate stages as $4/3<\gamma<5/3$, collapse before that the basic flow completes the condensation (i.e. $\gamma/2-5/6<0$) and the fragmentation of the background fluid is favored. On the other hand, if the gas cloud behaves adiabatically (i.e. $\gamma=5/3$), perturbations remain of the same order as the basic flow density \reff{rho-t-basic-flow}. We can conclude that, in this adiabatic-like case, pressure forces become progressively strong during the collapse as $\gamma$ increases having a stabilizing effect which prevents that density perturbations grow in amplitude with respect to the unperturbed flow. An intermediate type of behavior exists for $\gamma=4/3$, in this case the disturbances grow like $(-t)^{-13/6}$.

\section{Motion equations of an unperturbed viscous fluid}\label{Sec.3}
In this Section we discuss a model in order to build the motion equations of a spherically symmetric and uniform gas cloud including the corrections due to the presence of dissipative processes; the hypothesis that the fluid is initially at rest already stands here. The Lagrangian equation \reff{eq-lagrangian-0} describes a spherical shell which collapses under the gravitational action. In such an approach the shell results comoving with the collapsing background, this implies that there are no displacements between parts of fluid with respect to ones other since we assume an homogeneous and isotropic flow.

Dissipative processes are therefore related to the only fluid compression and can be phenomenologically described \cite{ll-fluid} by the presence of \emph{bulk viscosity} (or second viscosity) effects summarized by the corresponding coefficient $\zeta$. Furthermore, in this model we are able to neglect the shear viscosity (first viscosity) since it is connected with processes of relative motion among different parts of the fluid.

The effect of \emph{bulk viscosity} on the fluid motion can be expressed by the generation of a negative pressure additional to the thermostatic one. In the relativistic limit, dissipative precesses are derived \cite{ll-fluid}, \cite{e40}, \cite{clw92}, \cite{mont01} as a correction of the energy-momentum tensor pressure characterizing the matter filling the space-time. In presence of viscosity, the thermostatic pressure $p$ is replaced by a new quantity of the form
\begin{equation}
\label{p-tilde}
\tilde{p}=p-\zeta\,u^{\mu}_{;\,\mu}\;,
\end{equation}
where $u_\mu$ is the shell comoving 4-velocity, i.e. in the present case $u_\mu=(1,\textbf{0})$. The coefficient $\zeta$ is not constant and we have to express its dependence on the state parameters of the fluid. In the homogeneous model this quantity depends only on time and therefore we may consider it as a function of the fluid density $\rho$. According to literature developments \cite{bk76}, \cite{bk77}, \cite{bnk79}, \cite{m95} we assume that $\zeta$ depends on $\rho$ via a power-law of the form
\begin{equation}
\label{bulk}
\zeta=\zeta_0\,\rho^\nu\;,
\end{equation}
where $\zeta_0$ is a constant and $\nu$ is an adimensional parameter. 

Let us now evaluate equation \reff{p-tilde} in correspondence to the Newtonian limit appropriate to our analysis. As in Sec. \ref{Sec.2} , we consider here that gravitational forces are much greater than ones due to the thermostatic pressure in order to neglect $p$ in the zero-order analysis. In the non-relativistic approach, the metric we consider is a flat Minkowskian one expressed in spherical coordinates, i.e.
\begin{equation}
\label{flat-metric}
ds^{2}=dt^{2}-dr^{2}-r^{2}d\theta^{2}+r^{2}\sin^{2}\theta\,d\varphi^{2}\;,
\end{equation}
and the metric determinant is $g=-r^{4}\sin^{2}\theta$. For the 4-divergence presents in \reff{p-tilde} we immediately obtain, in this case, $u^{\mu}_{;\,\mu}=2\dot{r}/r\;$. 

Considering the basic flow density as $\rho=M/(\textnormal{\tiny{$\frac{4}{3}$}}\pi r^{3})$, the viscous pressure $\tilde{p}$ now reads
\begin{equation}
\tilde{p}= -2\zeta_0\left(3M/4\pi\right)^{\nu}\frac{\p r}{\p t}\;r^{-1-3\nu}\;.	
\end{equation}
The pressure force acting on the collapsing shell (and damping its motion) takes the form $F_{\tilde{p}}=\tilde{p}\,4\pi r^{2}$. By these considerations, the motion equation \reff{eq-lagrangian-0} for a viscous fluid becomes
\begin{equation}
\label{eq-lagr}
\frac{\partial^2r}{\partial t^2}=-\frac{GM}{r^2}-\frac{C}{r^{3\nu-1}}\,\frac{\p r}{\p t}\;,
\end{equation}
where $C=8\pi\zeta_0\left(3M/4\pi\right)^{\nu}$.

The above equation must be integrated to obtain the evolution of the radial velocity $v$ and the density $\rho$ of the unperturbed flow. In order to compare our viscous analysis with the Hunter case discussed in the previous Section, let us now require that the viscosity does not influence the final form of the velocity and, for instance, it should be yet proportional to $1/\sqrt{r}$ (see \reff{vel-v-0}). Substituting an expression of the form 
\begin{equation}
\label{vel-v}
v=B/\sqrt{r}
\end{equation}
into the equation \reff{eq-lagr} we see that, in correspondence to the choice $\nu=5/6$, it is again a solution as soon as we take the following identification
\begin{equation}
\label{B-coeff}
B=C-\sqrt{C^{2}+2GM}\;,\qquad\nu=5/6\;,
\end{equation}
where $B$ assumes only negative values. Although this dynamics is analytically integrable only for the particular value $\nu=5/6$, the obtained behavior $v\sim r^{-1/2}$ remains asymptotically (as $r\rightarrow0$) valid if the condition $\nu<5/6$ is satisfied. 

Using such a solution we are able to build an explicit form of the quantity $\beta$ defined by \reff{r-a}; differentiating this relation with respect to time and taking into account \reff{vel-v} we obtain a differential equation for the variable $\beta$ which admits the solution
\begin{equation}
\label{beta-t}
\cos\beta^{3}=3A\,(-t)\;,
\end{equation}
where $A$ is defined to be $A=-B/2a^{3/2}$. The Eulerian expressions \reff{basic-flow} of the unperturbed quantities $\textbf{v}$, $\rho$ and $\phi$ hold here since they are derived simply from relations \reff{r-a} and \reff{mass-conservation}; the effects of \emph{bulk viscosity} in this zero order analysis are now only summarized by the new time dependence \reff{beta-t} of the parameter $\beta$ which implies a different dynamics of the basic flow.

\section{Perturbations theory and the dynamics of fluctuations}\label{Sec.4}
In the last Section we have discussed the motion of a viscous basic flow which collapses under the action of its own gravitation. We shall now suppose that small disturbances appear on this field; the perturbations evolution can be described by the Eulerian equations \reff{eq-unpert-0} modified by the presence of dissipative processes (as in Sec. \ref{Sec.2} we consider here that the thermostatic pressure influences the first-order perturbations dynamics). Such kind of effects come out from the thermodynamical irreversibility of the collapse process and are due to the miscrophysics of non-equilibrium \cite{ll-fluid}.

Dissipation effects do not influence the continuity equation \reff{continuity-unpert-0} since it is built starting from the mass conservation law of the fluid; the Euler equation, on the other hand, requires to be modified by the presence of viscosity which causes irreversible transfer of momentum from points where the velocity is large to those ones where it is small. A derivation of these corrections is developed in \cite{ll-fluid} starting from general considerations about mechanical properties of the fluid. The Euler equation in presence of viscosity corresponds to the Navier-Stokes equation when a gravitational potential $\phi$ is included, i.e.
\begin{equation}
\label{eulero-unpert}
\dot{\textbf{v}}+(\textbf{v}\cdot\nabla)\cdot\textbf{v}=
-\nabla\phi-\frac{1}{\rho}\,\nabla p+\frac{\zeta}{\rho}\,\nabla(\nabla\cdot\textbf{v})\;,	
\end{equation}
where $\zeta$ is the \emph{bulk viscosity} coefficient \reff{bulk} (in this formula we have neglected the shear viscosity since the homogeneity hypothesis already holds). Beside this equation, the viscous fluid motion is described by the continuity equation \reff{continuity-unpert-0} and the Poisson one \reff{poisson-unpert-0}.

Once built motion equations, we start from the unperturbed solutions \reff{basic-flow} (which still hold here as seen in Sec. \ref{Sec.3}) to analyze the behavior, in presence of viscosity, of first-order perturbations to the basic flow. Let us now substitute expressions $(\textbf{v}+\delta\textbf{v})$, $(\rho+\delta\rho)$, $(\phi+\delta\phi)$ and $p+\delta p$ into the viscous Euler equation \reff{eulero-unpert} to obtain
\begin{equation}
\label{euler-pert}
\dot{\delta\textbf{v}}+\nabla(\textbf{v}\cdot\delta\textbf{v})+(\nabla\times\delta\textbf{v})\times\textbf{v}=
-\nabla\delta\phi\,-\,\frac{v_s^2}{\rho}\nabla\delta\rho\,+\,\frac{\zeta}{\rho}\,\nabla(\nabla\cdot\delta\textbf{v})\;.
\end{equation}
Proceeding as in the non-viscous case (Sec. \ref{Sec.2}) we shall now apply the \textbf{rot} operator to this last equation in order to get first the solution for the vorticity $\delta\textbf{w}$ and then the expression for the velocity perturbations $\delta\textbf{v}$. Indeed, using the vectorial identity $\textbf{rot}[\nabla f]=0$ (which holds for each scalar function $f$), all terms in the right hand side of \reff{euler-pert} vanish under this operation. In particular the term due to the viscous correction disappears from this equation because $\zeta$ is, by assumption, a space-independent function. In this way we reach the equation \reff{eq-vorticity} for the vorticity which yields the same solution \reff{dv} as in the non-viscous case. 

Following the line of the Hunter work we now build the equations for the perturbed quantities $\delta\textbf{v}$, $\delta\rho$ and $\delta\phi$. Substituting the expression \reff{dv} into the first-order perturbed Eulerian motion \reff{euler-pert}, we obtain (using \reff{r-a} and the conformal spherical coordinates $[a, \theta, \varphi]$), the equation
\begin{equation}
\label{euler-pert-2}
\dot{\psi}+\delta\phi+\frac{v_s^2}{\rho_0}\cos^6\beta\,\delta\rho-\frac{\zeta}{\rho_0}\cos^{2}D^{2}\psi=0\;,
\end{equation}
which corresponds to the viscous generalization of \reff{pert-2-0}. The other perturbed equations maintain their own forms \reff{pert-1-0} and \reff{pert-3-0} also in the viscous case.

Our analysis proceeds in order to build an unique equation which describes the evolution of the density perturbations. By following the procedure developed in the non-dissipative approach we get now the equation
\begin{equation}
\begin{aligned}
\label{eq-fundamental}
&\tfrac{\p}{\p t}\Big(\cos^{10}\beta\;\dot{\delta\rho}-6\;\sin\beta\,\cos^9\beta\,\dot{\beta}\,\delta\rho\Big)-4\pi G\rho_{0}\,\cos^4\beta\,\delta\rho=\\
&\qquad\quad=\Big({v_s}^2\,\cos^6\beta-6\,\frac{\zeta}{\rho_0}\,\sin\beta\cos^{11}\beta\,\dot{\beta}\Big)\;D^2\delta\rho+\frac{\zeta}{\rho_0}\,\cos^{12}\beta \;D^{2}\,\dot{\delta\rho}\;,
\end{aligned}
\end{equation}
that is the second-order dynamics which governs the density perturbations in the viscous regime.

\section{The asymptotic behavior of density perturbations}\label{Sec.5}
Following the non-viscous approach, let us now factorize perturbations $\delta\rho$ in plane waves by the formula \reff{fourier-trasform} and then use the barotropic relation $p=\kappa\rho^{\gamma}$, in which the limit cases $\gamma=1$ and $\gamma=5/3$ correspond respectively to an isothermal and an adiabatic behavior of the gas. According to these assumptions we are able to write the asymptotic form of equation \reff{eq-fundamental} near the end of the collapse as $(-t)\rightarrow0$.In this case, the quantity $\cos\beta$ is given by \reff{beta-t}, i.e.
\begin{equation}
\cos\beta=(3A)^{1/3}\;(-t)^{1/3},	
\end{equation}
and asymptotically we can make the approximation $\sin\beta\approx1$ in order to obtain an equation which generalizes \reff{eq-asym-0} in presence of viscosity; furthermore substituting expression \reff{rho-basic-flow} of the basic flow density $\rho$ into \reff{bulk}, we obtain
\begin{equation}
\label{z-beta)}
\zeta\,=\,\zeta_0\rho_0^{\nu}\,\,\cos^{-6\nu}\beta\;.
\end{equation}
With this assumptions, equation \reff{eq-fundamental} now reads 
\begin{equation}
\begin{aligned}
\label{eq-asym-nu}
&(-t)^2\,\ddot{\delta\varrho}-\bigg[\frac{16}{3}\,(-t)-\frac{\lambda}{(3A)^{2\nu-2/3}}(-t)^{8/3-2\nu}\bigg]\,\dot{\delta\varrho}\,+\\
&+\bigg[\frac{14}{3}-\frac{4\pi G \rho_0}{9 A^{2}}+\frac{v_0^2\, k^2}{(3A)^{2\gamma-2/3}}(-t)^{8/3-2\gamma}-\frac{2\lambda}{(3A)^{2\nu-2/3}}(-t)^{5/3-2\nu}\bigg] \delta\varrho=0\;,
\end{aligned}
\end{equation}
where $v_0^{2}=\kappa\gamma\rho_0^{\gamma-1}$ and the parameter $\lambda$ is given by
\begin{equation}
\label{lambda}
\lambda\,=\,\zeta_0\,\rho_0^{\,\nu-1}\,k^{2}\;.
\end{equation}

The background motion equations were derived in Sec. \ref{Sec.3} for a particular value of the \emph{bulk viscosity} parameter $\nu=5/6$. In this case, equation \reff{eq-asym-nu} assumes the form
\begin{equation}
\label{eq-asym}
(-t)^2\,\ddot{\delta\varrho}-\bigg[\frac{16}{3}-\frac{\lambda}{3A}\bigg](-t)\,\dot{\delta\varrho}+
\bigg[\frac{14}{3}-\frac{4\pi G \rho_0}{9 A^{2}}+\frac{v_0^2\, k^2(-t)^{8/3-2\gamma}}{(3A)^{2\gamma-2/3}}-\frac{2\lambda}{3A}\bigg] \delta\varrho=0\;.
\end{equation}

In analogy with the non-dissipative case, a complete solution of \reff{eq-asym} involves Bessel functions of first and second species $J$ and $Y$ respectively and it explicitly reads
\begin{equation}
\label{drho-solution}
\delta\varrho=\,C_1\,G_1(t)\,+\,C_2\,G_2(t)\;,
\end{equation}
where $C_1$, $C_2$ are integration constants and the functions $G_1$ and $G_2$ are defined to be
\begin{equation}
\label{g-functions}
G_1(t)=(-t)^{-\frac{13}{6}+\frac{\lambda}{6A}}\;J_n\big[q(-t)^{4/3-\gamma}\big]\;,\quad
G_2(t)=(-t)^{-\frac{13}{6}+\frac{\lambda}{6A}}\;Y_n\big[q(-t)^{4/3-\gamma}\big]\;,
\end{equation}
having set the Bessel parameters $n$ and $q$ as  
\begin{subequations}
\label{bessel-parameter}
\begin{align}
\label{parameter-n}
&n=[A^2-2\lambda A+\lambda^2+16\pi G \rho_0]^{\frac{1}{2}}\;/\,(6A(4/3-\gamma))\;,\\
\label{parameter-q}
&q=-k v_0(3A)^{1/3-\gamma}\;/\,(4/3-\gamma)\;.
\end{align}
\end{subequations}

We now proceed, in order to study the asymptotic evolution of the solution \reff{drho-solution}, analyzing the cases $1\leqslant\gamma<4/3$ and $4/3<\gamma\leqslant5/3$ separately using the asymptotic expansion for Bessel functions introduced in Sec. \ref{Sec.2}.

In the first case of an isothermal-like behavior of the gas, the time exponent inside Bessel functions results to be positive thus their argument is much less than unity. In this way, an asymptotic form of functions $G$ can be found as follow
\begin{equation}
\label{g-functions-iso}
G_{1}^{ISO}=c_{1}\;\,(-t)^{-\frac{13}{6}+\frac{\lambda}{6A}+(\frac{4}{3}-\gamma)n}\;,\qquad
G_{2}^{ISO}=c_{2}\;\,(-t)^{-\frac{13}{6}+\frac{\lambda}{6A}-(\frac{4}{3}-\gamma)n}\;,
\end{equation}
where $c_1$ and $c_2$ are constants quantities. The condition which implies the density perturbations collapse is that at least one of $G$ functions diverges as $(-t)\rightarrow0$. An analysis of time exponents yields that $G_1$ diverges if $\lambda<7A-2\pi G\rho_0/3A$ but, on the other hand, $G_2$ is always divergent for all $\lambda$. These results imply that, in the isothermal case, perturbations always condense. Let us now compare these collapses with the basic flow one; the background density evolves like $\cos^{-6}\beta$ (see \reff{rho-basic-flow}) that is, using \reff{beta-t}, in the same way of the non-viscous case, i.e.
\begin{equation}
\label{rho-t-basic-flow-visc}
\rho\sim(-t)^{-2}\;.
\end{equation}
In the non-dissipative case we have seen that perturbations grow more rapidly with respect to the background density involving the fragmentation of the basic flow independently on the value of $\gamma$; in presence of viscosity density contrasts assume the asymptotic form
\begin{equation}
\label{delta-iso}
\delta^{\,ISO}\sim(-t)^{-\frac{1}{6}+\frac{\lambda}{6A}-\frac{1}{6A}\sqrt{}[A^2-2\lambda A+\lambda^2+16\pi G \rho_0]}\;.
\end{equation}
Here the exponent is always negative and it does not depend on $\gamma$, this implies that $\delta^{\,ISO}$ diverges as the singularity is approached and real sub-structures are formed involving the basic flow fragmentation. This issue means that the viscous forces do not have enough strength to contrast an isothermal perturbations collapse implying a formation of an unique structure. 

For $4/3<\gamma\leqslant5/3$ the argument of Bessel functions becomes much gather than unity and $J$, $Y$ assume an oscillating behavior. In this adiabatic-like approach functions $G$ read
\begin{equation}
\label{g-functions-adb}
G_{1,2}^{\,ADB}=\tilde{c}_{1,2}\;\;^{\cos}_{\sin}\big[q(-t)^{4/3-\gamma}\big]\;(-t)^{\frac{\gamma}{2}-\frac{17}{6}+\frac{\lambda}{6A}}\;,	
\end{equation}
where $\tilde{c}_{1,2}$ are constants. Following the isothermal approach, we shall now analyze the time power-law exponent in order to determine the collapse conditions. $G$ functions diverge, involving perturbations condensation, if the parameter $\lambda$ is less than a threshold value: this condition reads $\lambda<17A-3A\gamma$ for a given value of the barotropic index $\gamma$. Expressing $\lambda$ in function of the wave number using expression \reff{lambda} with $\nu=5/6$, we outline, for a fixed viscous parameter $\zeta_0$, a constraint on $k$ which is similar to the condition appearing in the Jeans model \cite{j902}, \cite{j28}. The threshold value for the wave number is
\begin{equation}
\label{k-critic}
k_C=\sqrt{(17A-3\gamma A)\rho_0^{1/6}\,/\,\zeta_0}
\end{equation}
and therefore the condition for the density perturbations collapse, i.e. $\delta\rho^{ADB}\rightarrow\infty$, reads $k<k_C$, recalling that, in the Jeans model for a static background, the condition for the collapse $k<k_J=[4\pi G\rho_0\,/\,v_s^{2}\,]^{\frac{1}{2}}$ holds. It is to be remarked that, in absence of viscosity ($\zeta_0=0$), the expression \reff{k-critic} diverges implying that all perturbations scales can be conducted to the collapse as in the non-dissipative adiabatic-like case. On the other hand, if we consider perturbations of fixed wave number, they asymptotically decrease as $(-t)\rightarrow0$ for $\lambda>17A-3A\gamma$. Thus for each $k$ there is a value of the \emph{bulk viscosity} coefficient over which the dissipative forces contrast the formation of sub-structures.

If $k<k_C$, perturbations oscillate with ever increasing frequency and amplitude. For a non-zero viscosity coefficient, density contrasts evolve like
\begin{equation}
\label{drho-adb}
\delta^{\,ADB}\,\sim\,(-t)^{\frac{\gamma}{2}-\frac{5}{6}+\frac{\lambda}{6A}}\;.
\end{equation}
A study of the time exponent yields the introductions of a new threshold value. If $\lambda<5A-3A\gamma$, i.e. the viscosity is enough small, sub-structures form; on the other hand, when the parameter $\zeta_0$, or the wave number $k$, provides a $\lambda$-term overcoming this value, the perturbations collapse is so much contrasted that no fragmentation process occurs. In other words, if $\lambda>5A-3A\gamma$ we get $\delta^{\,ADB}\rightarrow0$, i.e. for a given $\gamma$ there is a viscous coefficient $\zeta_0$ enough large ables to prevents the sub-structures formation. It is remarkable that in the adiabatic case, $\gamma=5/3$, dissipative processes, of any magnitude order, contrast the fragmentation because, while the Jeans-like length survives, the threshold value for sub-structures formation approaches infinity. We can conclude that, in the adiabatic-like case for $4/3<\gamma\leqslant5/3$, the fragmentation in the \emph{top-down} scheme is deeply unfavored by the presence of \emph{bulk viscosity} which strongly contrasts the density perturbations collapse.

To complete this Section we point out two relevant questions about the generality and the validity of our zero and first-order analysis.\vspace{0.5cm}

(i) \emph{Physical meaning of $\nu=5/6$} --- We now clarify why the choice $\nu=5/6$ is appropriate to a consistent treatment of the asymptotic viscous collapse. We start by observing that bulk viscous effects can be treated in a predictive way only if they behave as small corrections to the thermodynamical system. To this respect we have to require that the asymptotic collapse is yet appropriately described by the non viscous background flow. As soon as we recognize that equation \reff{eq-lagr} can be rewritten as follows
\begin{equation}
v\frac{\p v}{\p r}+\frac{C}{r^{3\nu-1}}\,v=-\frac{GM}{r^{2}}\;,
\end{equation}
it is easy to infer that in the asymptotic limit as $r\rightarrow0$, the non viscous behavior $v\sim r^{-1/2}$ is preserved only if $\nu\leqslant5/6$; in fact, in correspondence to this restriction, the viscous correction, behaving like $\mathcal{O}(r^{-3\nu+1/2})$, is negligible with respect to the leading order $\mathcal{O}(r^{-2})$ when the singularity is approached; therefore the request that the viscosity is a small correction implies the choice $\nu\leqslant5/6$.

On the other hand, if we take $\nu=5/6-\Delta$, with $\Delta>0$, then the perturbations dynamics in the viscous case \reff{eq-asym-nu} rewrites as
\begin{equation}
\begin{aligned}
\label{eq-asym-nu-delta}
&(-t)^2\,\ddot{\delta\varrho}-\bigg[\frac{16}{3}-\frac{\lambda}{(3A)^{1-2\Delta}}(-t)^{2\Delta}\bigg]\, (-t)\,\dot{\delta\varrho}\,+\\
&+\bigg[\frac{14}{3}-\frac{4\pi G \rho_0}{9 A^{2}}+\frac{v_0^2\, k^2}{(3A)^{2\gamma-2/3}}(-t)^{8/3-2\gamma}-\frac{2\lambda}{(3A)^{1-2\Delta}}(-t)^{2\Delta}\bigg] \delta\varrho=0\;.
\end{aligned}
\end{equation}
As far as we deal with the adiabatic-like case, it is immediate to verify that, as $(-t)\rightarrow0$, the viscous terms in $\dot{\delta\varrho}$ and in $\delta\varrho$ respectively are negligible and the dynamics matches asymptotically the non-viscous Hunter result \reff{delta-adb-0} (apart from non-relevant features). In the isothermal-like case the viscous term in $\dot{\delta\varrho}$ is again negligible, but for $\Delta<4/3-\gamma$ the one appearing in $\delta\varrho$ could now dominate the pressure one; however as $(-t)\rightarrow0$ both these terms provide higher order corrections with respect to the constants in $\delta\varrho$ and equation \reff{eq-asym-nu-delta} reduces to an equation whose solution overlaps the Hunter behavior \reff{delta-iso-0} (we remark that for $\nu<5/6$ the viscous parameters asymptotically disappears from the background dynamics too).     

Matching together the above considerations for the zero and first-order re-spectively, we see that $\nu=5/6$ is the only physical value which does not affect the background dynamics but makes important the viscous corrections in the asymptotic behavior of the density contrasts.\vspace{0.5cm}

(ii) \emph{Validity of the Newtonian approximation} --- Since our analysis addresses Newtonian dynamics while the cloud approaches the extreme collapse, it is relevant to precise the conditions which ensure the validity of such a scheme. The request that the shell corresponding to the radial coordinate $r$ lives in the Newtonian paradigm leads to impose that it remains greater than its own \emph{Schwarzschild Radius}, i.e. 
\begin{equation}
r(t)\gg 2GM(a)\;,
\end{equation}
where $M(a)=\rho_0\,(\textnormal{\tiny{$\frac{4}{3}$}}\pi a^3)$ and by \reff{r-a} together with the solution \reff{beta-t} we reache the inequality
\begin{equation}
\label{t-in}
(-t)\gg-\textnormal{\tiny{$\frac{2}{3}$}}\left(\textnormal{\tiny{$\frac{8}{3}$}}\pi G \rho_0\right)^{3/2}
\left[8\pi \zeta_0\rho_0^\nu-\sqrt{(8\pi \zeta_0\rho_0^\nu)^2+\textnormal{\tiny{$\frac{8}{3}$}}\pi G \rho_0 \,a^{3-6\nu}}\;\,\right]^{-1}\,a^{9/2-3\nu}\;.
\end{equation}
Once fixed the fundamental parameters $a$, $\rho_0$ and $\zeta_0$, the above constraint on the time variable states up to which limit a shell remains appropriately described by the Newtonian approach.

About the dynamics of a physical perturbations scale $l=(2\pi/k)\cos\beta^{2}$ (here $\cos\beta^{2}$ plays the same role of a cosmic scale factor), its Newtonian evolution is ensured by the linear behavior, as soon as, condition \reff{t-in} for the background holds. More precisely a perturbations scale is Newtonian if its size is much smaller than the typical space-time curvature length, but for a weak gravitational field this requirement must have no physical relevance. To explicit such a condition, we require that the physical perturbations scale is much greater than its own \emph{Schwarzschild Radius}, which leads to the inequality $k\gg\chi(-t)^{-1/3}$, where $\chi=\left[\textnormal{\tiny{$\frac{4}{3}$}}(2\pi)^{3}G\rho_0(3A)^{-2/3}\right]^{1/2}$; combining this result with the inequality \reff{t-in} we arrive to the following constraint
\begin{equation}
\label{k-t-lim}
k\gg\frac{2\pi}{(3A)^{8/3}\;a}\;,
\end{equation}
being $A=-\textnormal{\tiny{$\frac{1}{2}$}}\,a^{3\nu-3/2}\left[8\pi \zeta_0\rho_0^\nu-\sqrt{(8\pi \zeta_0\rho_0^\nu)^2+\textnormal{\tiny{$\frac{8}{3}$}}\pi G \rho_0 \,a^{3-6\nu}}\;\,\right]$. The condition \reff{k-t-lim} tells us which modes are Newtonian within the shell whose initial radius takes the value $a$.

\section{Concluding remarks}\label{Sec.6}
Our analysis outlined how the presence of \emph{bulk viscosity} induces a deep modification of the extreme gravitational collapse relative to an uniform, spherically symmetric and dust-like gas cloud. While the isothermal-like collapse ($1\leqslant\gamma<4/3$) is characterized by sub-structures formation even when viscous effects are taken into account, the adiabatic-like one ($4/3<\gamma\leqslant5/3$) undergoes an opposite asymptotic regime as soon as the viscosity become sufficiently intense. Though \emph{bulk viscosity} does not affect (by hypothesis) the extreme collapse of the background flow, nevertheless its presence changes drastically the dynamics of perturbations which are damped at the point to generate vanishing density contrasts. Thus, in the adiabatic case, the fate of a collapsing cloud is sensitive to the viscous effects by itself induced. In particular \emph{bulk viscosity} is able to restore a kind of Jeans length for the cloud perturbations; scales above this threshold begin to collapse but, if below the second threshold, no sub-structures formation takes place.

The interest in the model here presented comes out because it is expectable that extreme regimes of collapse are associated with \emph{bulk viscosity} which is a macroscopic effect of the non-equilibrium thermodynamics characterizing microphysical precesses. However this study has to be upgraded by including relativistic effects which becomes asymptotically relevant for the small scales. The idea is to restate the problem in terms of a system dynamics based on the Einstein equations adapted to the present context. In this respect it would be worth to consider radial inhomogeneities of the gas cloud and to include thermostatic pressure contributions already on the background flow evolution. Such a generalized scheme would allow to evaluate how important \emph{bulk viscosity} is in favoring black holes or compact astrophysical objects with respect to the fragmentation process in sub-structures. We will address such line of thinking, relevant for cosmological and astrophysical topics, as subject for further investigations in the \emph{bulk viscosity} phenomenology.

\end{document}